\documentclass[prd,superscriptaddress,unsortedaddress,twocolumn,showpacs,preprintnumbers,amsmath,amssymb]{revtex4}
\usepackage[dvips]{graphicx}
\usepackage{amsmath,amssymb,times}
\usepackage{ulem}

\newcommand{\beq}{\begin{equation}}
\newcommand{\eeq}{\end{equation}}
\newcommand{\bea}{\begin{eqnarray}}
\newcommand{\eea}{\end{eqnarray}}

\renewcommand{\b}{\beta}

\DeclareSymbolFont{boldletters}{OML}{cmm} {b}{it}
\DeclareSymbolFontAlphabet{\mathbit}{boldletters}
\DeclareMathSymbol{\alpha}{\mathalpha}{letters}{"0B}
\DeclareMathSymbol{\beta}{\mathalpha}{letters}{"0C}
\DeclareMathSymbol{\gamma}{\mathalpha}{letters}{"0D}
\DeclareMathSymbol{\delta}{\mathalpha}{letters}{"0E}
\DeclareMathSymbol{\epsilon}{\mathalpha}{letters}{"0F}
\DeclareMathSymbol{\zeta}{\mathalpha}{letters}{"10}
\DeclareMathSymbol{\eta}{\mathalpha}{letters}{"11}
\DeclareMathSymbol{\theta}{\mathalpha}{letters}{"12}
\DeclareMathSymbol{\iota}{\mathalpha}{letters}{"13}
\DeclareMathSymbol{\kappa}{\mathalpha}{letters}{"14}
\DeclareMathSymbol{\lambda}{\mathalpha}{letters}{"15}
\DeclareMathSymbol{\mu}{\mathalpha}{letters}{"16}
\DeclareMathSymbol{\nu}{\mathalpha}{letters}{"17}
\DeclareMathSymbol{\xi}{\mathalpha}{letters}{"18}
\DeclareMathSymbol{\pi}{\mathalpha}{letters}{"19}
\DeclareMathSymbol{\rho}{\mathalpha}{letters}{"1A}
\DeclareMathSymbol{\sigma}{\mathalpha}{letters}{"1B}
\DeclareMathSymbol{\tau}{\mathalpha}{letters}{"1C}
\DeclareMathSymbol{\upsilon}{\mathalpha}{letters}{"1D}
\DeclareMathSymbol{\phi}{\mathalpha}{letters}{"1E}
\DeclareMathSymbol{\chi}{\mathalpha}{letters}{"1F}
\DeclareMathSymbol{\psi}{\mathalpha}{letters}{"20}
\DeclareMathSymbol{\omega}{\mathalpha}{letters}{"21}
\DeclareMathSymbol{\varepsilon}{\mathalpha}{letters}{"22}
\DeclareMathSymbol{\vartheta}{\mathalpha}{letters}{"23}
\DeclareMathSymbol{\varpi}{\mathalpha}{letters}{"24}
\DeclareMathSymbol{\varrho}{\mathalpha}{letters}{"25}
\DeclareMathSymbol{\varsigma}{\mathalpha}{letters}{"26}
\DeclareMathSymbol{\varphi}{\mathalpha}{letters}{"27}
\DeclareMathSymbol{\Gamma}{\mathalpha}{letters}{"00}
\DeclareMathSymbol{\Delta}{\mathalpha}{letters}{"01}
\DeclareMathSymbol{\Theta}{\mathalpha}{letters}{"02}
\DeclareMathSymbol{\Lambda}{\mathalpha}{letters}{"03}
\DeclareMathSymbol{\Xi}{\mathalpha}{letters}{"04}
\DeclareMathSymbol{\Pi}{\mathalpha}{letters}{"05}
\DeclareMathSymbol{\Sigma}{\mathalpha}{letters}{"06}
\DeclareMathSymbol{\Upsilon}{\mathalpha}{letters}{"07}
\DeclareMathSymbol{\Phi}{\mathalpha}{letters}{"08}
\DeclareMathSymbol{\Psi}{\mathalpha}{letters}{"09}
\DeclareMathSymbol{\Omega}{\mathalpha}{letters}{"0A}

\begin{document}
\preprint{SAGA-HE-238-07}
\title{
Polyakov loop extended NJL model with imaginary chemical potential}

\author{Yuji Sakai}
\email[]{sakai2scp@mbox.nc.kyushu-u.ac.jp}
\affiliation{Department of Physics, Graduate School of Sciences, Kyushu University,
             Fukuoka 812-8581, Japan}
\author{Kouji Kashiwa}
\email[]{kashiwa2scp@mbox.nc.kyushu-u.ac.jp}
\affiliation{Department of Physics, Graduate School of Sciences, Kyushu University,
             Fukuoka 812-8581, Japan}

\author{Hiroaki Kouno}
\email[]{kounoh@cc.saga-u.ac.jp}
\affiliation{Department of Physics, Saga University,
             Saga 840-8502, Japan}

\author{Masanobu Yahiro}
\email[]{yahiro2scp@mbox.nc.kyushu-u.ac.jp}
\affiliation{Department of Physics, Graduate School of Sciences, Kyushu University,
             Fukuoka 812-8581, Japan}

\date{\today}

\begin{abstract}
The Polyakov loop extended Nambu--Jona-Lasinio 
(PNJL) model with imaginary chemical potential 
is studied. The model possesses the extended ${\mathbb Z}_{3}$ 
symmetry that QCD does. 
Quantities invariant under the extended ${\mathbb Z}_{3}$ 
symmetry, such as the partition function, 
the chiral condensate and the modified Polyakov loop, 
have the Roberge-Weiss (RW) periodicity. 
The phase diagram of confinement/deconfinement transition 
derived with the PNJL model is 
consistent with 
the RW prediction on it and the results of lattice QCD. 
The phase diagram of chiral transition is also 
presented by the PNJL model. 
\end{abstract}

\pacs{11.30.Rd, 12.40.-y}
\maketitle


With the aid of the progress in computer power, 
lattice QCD simulations have become feasible for thermal systems 
at zero quark chemical potential ($\mu$)~\cite{Kog}. 
As for $\mu>0$, however, 
lattice QCD has the well-known sign problem, and then 
the results are still far from perfection; for example, see 
Ref.~\cite{Kogut2} and references therein. 

Several approaches have been proposed to solve the sign problem. 
One of them is the use of imaginary chemical potential, 
since the fermionic determinant that appears in the euclidean partition 
function is real in the case; for example, see 
Refs.~\cite{FP1,FP2,Elia,Elia2,Chen,Lomb} and references therein.  
If the physical quantity such as chiral condensate is known in 
the imaginary $\mu$ region, one can extrapolate it to the real $\mu$ 
region, until there appears a discontinuity. 
Furthermore, in principle, one can evaluate with the Fourier transformation 
the canonical partition function with fixed quark number 
from the grand canonical partition function with imaginary 
chemical potential~\cite{RW}.

Roberge and Weiss (RW)~\cite{RW} found that the partition function of 
SU($N$) gauge theory with imaginary 
chemical potential $\mu=i\theta/\beta$,  
\begin{align}
Z(\theta )& = \int D\psi D\bar{\psi} DA_\mu 
\exp 
\Big[ 
- \int d^{4}x
\nonumber \\
&
\big\{ 
\bar{\psi}(\gamma D-m_0)\psi 
-{\frac{1}{4}}F_{\mu\nu}^2 
-i{\theta\over{\beta}}\bar{\psi}\gamma_4\psi
\big\}
\Big] ,
\label{eq:EQ1}
\end{align}
is a periodic function of $\theta$ with a period $2\pi/N$, that is
$Z(\theta+2{\pi}k/N)=Z(\theta)$ for any integer $k$, 
by showing that $Z(\theta+2{\pi}k/N)$ is reduced to $Z(\theta)$ with 
the ${\mathbb Z}_{N}$ transformation 
\bea
\psi \to U \psi, \quad 
A_{\nu} \to UA_{\nu}U^{-1} - \tfrac{i}{g} (\partial_{\nu}U)U^{-1} \;,
\label{z3}
\eea
where $U(x,\tau)$ are elements of SU($N$) with the boundary condition 
$
U(x,\beta)=\exp(-2i \pi k/N)U(x,0).  
$
Here $\psi$ is the fermion field with mass $m_0$, 
$F_{\mu\nu}$ is the strength tensor of the 
gauge field $A_\mu$, and $\beta$ is the inverse of temperature $T$. 
The RW periodicity means that 
$Z(\theta)$ is invariant under the 
extended ${\mathbb Z}_{N}$ transformation 
\bea
\theta \to \theta +  \tfrac{2 \pi k}{N}, \
\psi \to U \psi, \
A_{\nu} \to UA_{\nu}U^{-1} - \tfrac{i}{g} (\partial_{\nu}U)U^{-1}. 
\label{extended-z3}
\eea
Quantities invariant under the extended ${\mathbb Z}_N$ 
transformation, such as the thermodynamic potential $\Omega(\theta)$ and 
the chiral condensate, keep the RW periodicity. 
Meanwhile, the Polyakov loop $\Phi$ is not invariant under 
the transformation (\ref{extended-z3}), since 
it is transformed as 
$\Phi \to \Phi e^{-i{2\pi k/N}}$. 
In general, non-invariant quantities 
such as $\Phi$  do not have the periodicity. 
Roberge and Weiss also showed with perturbation that in the high $T$ region 
$d\Omega(\theta)/d\theta$ and $\Phi$ are discontinuous 
as a function of $\theta$ at values of ${(2k+1)\pi/N}$, and 
also found with the strong coupled lattice theory that 
the discontinuities disappear in the low $T$ region. 
This is called the Roberge-Weiss phase transition of first order, 
and is observed in lattice simulations~\cite{FP1,FP2,Elia,Elia2,Chen,Lomb}.

\begin{figure}[htbp]
\begin{center}
 \includegraphics[width=0.4\textwidth]{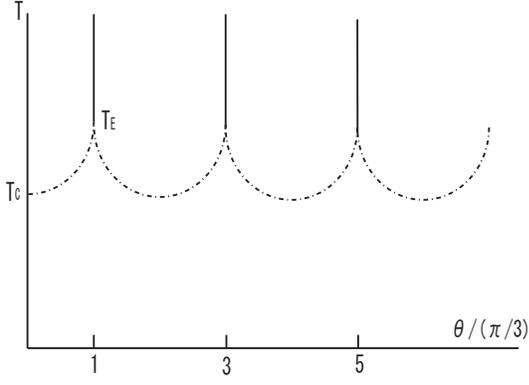} 
\end{center}
\caption{
The RW prediction on the QCD phase diagram in 
the $\theta-T$ plane. 
The solid lines represent the confinement/deconfinement phase 
transition of first order, 
and the dot-dashed ones correspond to 
the chiral phase transition of second order. 
}
\label{fig1}
\end{figure}

Figure \ref{fig1} shows a predicted phase diagram in the $\theta$-$T$ plane. 
The solid lines represent the RW discontinuities of 
the Polyakov loop, and the dot-dashed lines do 
the chiral phase transition predicted by the 
lattice simulations, although results of the simulations 
are not conclusive yet since it is hard to take the chiral limit 
in the simulations. 
In this paper, the term  ``chiral phase transition'' 
(``deconfinement phase transition'')
is used whenever the chiral condensate (the Polyakov loop) 
is discontinuous or not smooth. 

As an approach complementary to first-principle 
lattice simulations, one can consider several effective models. 
One of them is the Nambu--Jona-Lasinio (NJL) model~\cite{NJ1}.
Although the NJL model is a useful method for understanding the chiral symmetry breaking, this model does not possess a confinement mechanism.
As a reliable model that can treat both the chiral and the deconfinement phase transition,  we can consider the Polyakov loop extended NJL (PNJL) model
~\cite{Meisinger,Fukushima,Ghos,Megias,Ratti1,Ratti2,Rossner,Hansen,Sasaki1,Kashiwa3}.
In the PNJL model the deconfinement phase
transition is described by the Polyakov loop. 
It is known that effects of the Polyakov loop make the critical endpoint~\cite{AY,KKKN,Kashiwa1,Kashiwa2} move to higher $T$ and lower $\mu$ than the NJL model predicts~\cite{Rossner,Kashiwa3}. 

The PNJL model has the extended 
${\mathbb Z}_3$ symmetry needed to reproduce the RW periodicity, as shown 
later. 
In this paper, we study the phase diagram in the $\theta$-$T$ plane 
by using the PNJL model. Both the chiral and the deconfinement 
phase transition are analyzed in the chiral ($m_0=0$) limit 
where the lattice simulation is not available.

The model we consider is 
the following two-flavor PNJL Lagrangian,
\begin{align}
 {\cal L}  =& {\bar q}(i \gamma_\nu D^\nu -m_0)q \notag\\
             &\hspace{3mm} + G_{\rm s}[({\bar q}q)^2 
                          +({\bar q}i\gamma_5 {\vec \tau}q)^2] 
              - {\cal U}(\Phi [A],{\Phi} [A]^*,T) ,
             \label{eq:E1}
\end{align}
where $q$ denotes the two-flavor quark field, 
$m_0$ does the current quark mass, and 
$D^\nu=\partial^\nu-iA^\nu + \delta_{0}^{\nu}\mu$ 
for the chemical potential $\mu=i\theta/\beta$. 
The field $A^\nu$ is defined as 
$A^\nu=\delta_{\nu0}gA^\nu_a{\lambda^a\over{2}}$
with the gauge field $A^\nu_a$, 
the Gell-Mann matrix $\lambda_a$ and the gauge coupling $g$.
In the NJL sector, 
${\vec \tau}$ stands for the isospin matrix, and  
$G_{\rm s}$ denotes the coupling constant of the scalar type 
four-quark interaction. 
The Polyakov potential ${\cal U}$, defined later in ~\eqref{eq:E13} , 
is a function of the Polyakov loop $\Phi$ and its complex 
conjugate $\Phi^*$,
\begin{align}
\Phi      = {1\over{N_{\rm c}}}{\rm Tr} L,~~~~
\Phi^{*}  = {1\over{N_{\rm c}}} {\rm Tr}L^\dag ,
\end{align}
with
\begin{align}
L({\bf x})  = {\cal P} \exp\Bigl[
                {i\int^\beta_0 d \tau A_4({\bf x},\tau)}\Bigr],
\end{align}
where ${\cal P}$ is the path ordering, $A_4 = iA_0 $ and $N_c=3$. 
In the chiral limit ($m_0=0$), 
the Lagrangian density has the exact 
$SU(2)_{\rm L} \times SU(2)_{\rm R}
\times U(1)_{\rm v} \times SU(3)_{\rm c}$  symmetry. 

The temporal component $A_4$ is diagonal 
in the flavor space, because the color and the flavor space 
are completely separated out in the present case. 
In the Polyakov gauge, $L$ can be written in a diagonal form 
in the color space~\cite{Fukushima}: 
\begin{align}
L 
=  e^{i \beta (\phi_3 \lambda_3 + \phi_8 \lambda_8)}
= {\rm diag} (e^{i \beta \phi_a},e^{i \beta \phi_b},
e^{i \beta \phi_c} ),
\label{eq:E6}
\end{align}
where $\phi_a=\phi_3+\phi_8/\sqrt{3}$, $\phi_b=-\phi_3+\phi_8/\sqrt{3}$
and $\phi_c=-(\phi_a+\phi_b)=-2\phi_8/\sqrt{3}$. 
The Polyakov loop $\Phi$ is an exact order parameter of the spontaneous 
${\mathbb Z}_3$ symmetry breaking in the pure gauge theory.
Although the ${\mathbb Z}_3$ symmetry is not exact
in the system with dynamical quarks, it still seems to be a good indicator of 
the deconfinement phase transition as discussed later. 
Therefore, we use $\Phi$ to define the deconfinement phase transition.

Under the mean field approximation (MFA), the Lagrangian density becomes
\begin{align}
{\cal L}_{\rm MFA} =& 
{\bar q}( i \gamma_\mu D^\mu - (m_0+\Sigma_{\rm s}) )q \notag\\
&\hspace{10mm} - U(\sigma) - {\cal U}(\Phi,\Phi^{*},T), 
\label{eq:E7} 
\end{align}
where
\begin{align}
\sigma = \langle \bar{q}q \rangle, ~~~~~
\Sigma_{\rm s} = -2 G_{\rm s} \sigma , ~~~~~
U= G_{\rm s} \sigma^2. 
\label{eq:E10}
\end{align}
Using the usual techniques, 
one can obtain the thermodynamic potential
\begin{align}
\Omega =& -2 N_fV \int \frac{d^3{\rm p}}{(2\pi)^3}
         \Bigl[ 3 E ({\rm p}) \notag\\
        & \hspace{-4mm}+ \frac{1}{\beta}
           \ln~ [1 + 3(\Phi+\Phi^{*} e^{-\beta E^-({\bf p})}) 
           e^{-\beta E^-({\bf p})}+ e^{-3\beta E^-({\bf p})}]
         \notag\\
        & \hspace{-4mm}+ \frac{1}{\beta} 
           \ln~ [1 + 3(\Phi^{*}+{\Phi e^{-\beta E^+({\bf p})}}) 
              e^{-\beta E^+({\bf p})}+ e^{-3\beta E^+({\bf p})}]
	      \Bigl]\notag\\
        & \hspace{-4mm}+(U+{\cal U})V, 
\label{eq:E12} 
\end{align}
where $E({\rm p})=\sqrt{{\bf p}^2+M^2}$, 
$E^\pm({\rm p})=E({\rm p})\pm i\theta/\beta$ and $M=m_0 + \Sigma_{\rm s}$. 
We use ${\cal U}$ of Ref.~\cite{Ratti1} that is fitted to the result of 
lattice simulation in the pure gauge theory 
at finite $T$~\cite{Boyd,Kaczmarek}: 
\begin{align}
&{{\cal U}\over{T^4}} =  -\frac{b_2(T)}{2} {\Phi}^*\Phi
              -\frac{b_3}{6}({\Phi^*}^3+ \Phi^3)
              +\frac{b_4}{4}({\Phi}^*\Phi)^2, \label{eq:E13} \\
&b_2(T)   = a_0 + a_1\Bigl(\frac{T_0}{T}\Bigr)
                 + a_2\Bigl(\frac{T_0}{T}\Bigr)^2
                 + a_3\Bigl(\frac{T_0}{T}\Bigr)^3,  \label{eq:E14}
\end{align}
where parameters are summarized in Table I.  
The Polyakov potential yields a deconfinement phase transition at 
$T=T_0$ in the pure gauge theory.
Hence, $T_0$ is taken to be $270$ MeV predicted 
by the pure gauge lattice QCD calculation.

\begin{table}[h]
\begin{center}
\begin{tabular}{llllll}
\hline
~~~~~$a_0$~~~~~&~~~~~$a_1$~~~~~&~~~~~$a_2$~~~~~&~~~~~$a_3$~~~~~&~~~~~$b_3$~~~~~&~~~~~$b_4$~~~~~
\\
\hline
~~~~6.75 &~~~~-1.95 &~~~~2.625 &~~~~-7.44 &~~~~0.75 &~~~~7.5 
\\
\hline
\end{tabular}
\caption{
Summary of the parameter set in the Polyakov sector
used in Ref.~\cite{Ratti1}. 
All parameters are dimensionless. 
}
\end{center}
\end{table}

The variables of $\Phi$, ${\Phi}^*$ and $\sigma$ 
satisfy the stationary conditions, 
\bea
\partial \Omega/\partial \Phi=0, \quad
\partial \Omega/\partial \Phi^{*}=0, \quad
\partial \Omega/\partial \sigma=0 . 
\label{condition}
\eea
The thermodynamic potential $\Omega(\theta)$ at each $\theta$ is obtained by 
inserting the solutions, $\Phi(\theta)$, ${\Phi}(\theta)^*$ and  
$\sigma(\theta)$, of (\ref{condition}) at each $\theta$ into (\ref{eq:E12}). 

The thermodynamic potential $\Omega$ is not 
invariant under the ${\mathbb Z}_3$ transformation, 
$\Phi(\theta) \to \Phi(\theta) e^{-i{2\pi k/{3}}}$ and 
$\Phi(\theta)^{*} \to \Phi(\theta)^{*} e^{i{2\pi k\over{3}}}$, 
although 
${\cal U}$ of (\ref{eq:E13}) is invariant. 
Instead of the ${\mathbb Z}_3$ symmetry, however, 
$\Omega$ is invariant under the extended ${\mathbb Z}_3$ transformation, 
\begin{align}
&e^{\pm i \theta} \to e^{\pm i \theta} e^{\pm i{2\pi k\over{3}}},\quad  
\Phi(\theta)  \to \Phi(\theta) e^{-i{2\pi k\over{3}}}, 
\notag\\
&\Phi(\theta)^{*} \to \Phi(\theta)^{*} e^{i{2\pi k\over{3}}} .
\label{eq:K2}
\end{align}
It is convenient to introduce new variables 
$\Psi \equiv e^{i\theta}\Phi$ and 
$\Psi^{*} \equiv e^{-i\theta}\Phi^{*}$ 
invariant under the transformation (\ref{eq:K2}).
\begin{widetext} 
The extended ${\mathbb Z}_3$ transformation is then 
rewritten into 
\begin{align}
e^{\pm i \theta} \to e^{\pm i \theta} e^{\pm i{2\pi k\over{3}}}, \quad
\Psi(\theta) \to \Psi(\theta), \quad 
\Psi(\theta)^{*} \to \Psi(\theta)^{*} ,
\label{eq:K2'}
\end{align}
and $\Omega$ is also into  
\begin{align}
\Omega =  -2 N_fV \int \frac{d^3{\rm p}}{(2\pi)^3}
          \Bigl[ 3 E ({\rm p}) 
          &+ \frac{1}{\beta}\ln~ [1 + 3\Psi e^{-\beta E({\bf p})}
          + 3\Psi^{*}e^{-2\beta E({\bf p})}e^{\beta \mu_{\rm B}}
          + e^{-3\beta E({\bf p})}e^{\beta \mu_{\rm B}}]
\notag\\
          &\hspace{-5mm}+ \frac{1}{\beta} 
           \ln~ [1 + 3\Psi^{*} e^{-\beta E({\bf p})}
          + 3\Psi e^{-2\beta E({\bf p})}e^{-\beta\mu_{\rm B}}
          + e^{-3\beta E({\bf p})}e^{-\beta\mu_{\rm B}}]
	      \Bigl]+UV 
\notag\\
          &+\Bigl[-{b_2(T)T^4\over{2}}\Psi^{*} \Psi
          -{\b_3(T)T^4\over{6}}({\Psi^{*}}^3 e^{\beta \mu_{\rm B}}
          +\Psi^3 e^{-\beta\mu_{\rm B}})
          +{b_4T^4\over{4}}(\Psi^{*} \Psi)^2\Bigl]V ,
\label{eq:K3} 
\end{align}
\end{widetext}
where $\mu_{\rm B}=3 \mu= i 3 \theta/\beta$ 
is the baryonic chemical potential and 
the factor $e^{\pm\beta\mu_{\rm B}}$ is invariant 
under the transformation (\ref{eq:K2'}). 
Obviously, $\Omega$ is invariant under the transformation 
(\ref{eq:K2'}).

Under the transformation $\theta \to \theta + 2\pi k/3$, 
(\ref{eq:K3}) keeps the same form, 
if $\Psi(\theta)$ and $\Psi(\theta)^{*}$ are 
replaced by 
$\Psi(\theta +{2\pi k/3})$ and 
$\Psi(\theta +{2\pi k/3})^{*}$, respectively. 
This means that the stationary conditions for $\Psi(\theta)$ 
and $\Psi(\theta)^{*}$ 
agree with those for $\Psi(\theta +{2\pi k/3})$ and 
$\Psi(\theta +{2\pi k/3})^{*}$, respectively, and then that 
\begin{align}
\Psi(\theta +{\tfrac{2\pi k}{3}})=\Psi(\theta)
\ \ {\textrm{and}}
\ \  
\Psi(\theta +{\tfrac{2\pi k}{3}})^{*}=\Psi(\theta)^{*} . 
\label{psi-RW}
\end{align}

The potential $\Omega$ depends on $\theta$ through 
$\Psi(\theta)$, $\Psi(\theta)^{*}$, $\sigma(\theta)$ 
and $e^{i\theta}$. We then denote $\Omega(\theta)$ by  
$\Omega(\theta)=
\Omega(\Psi(\theta),\Psi(\theta)^{*},e^{i\theta})$, 
where $\sigma(\theta)$ is suppressed since it is irrelevant to 
proofs shown below. 
The RW periodicity of $\Omega$ is then shown as 
\begin{align}
\Omega (\theta + {2\pi k\over{3}})=&
\Omega ( \Psi(\theta),
\Psi(\theta)^{*}, 
e^{i{2\pi k\over{3}}+i \theta})
\notag\\
=&\Omega (\Psi(\theta),\Psi(\theta)^{*},
e^{i\theta})=\Omega(\theta), 
\label{eq:K4b}
\end{align}
by using (\ref{psi-RW}) in the 
first equality and the extended ${\mathbb Z}_3$ symmetry of $\Omega$ 
in the second equality.

Equation (\ref{eq:K3}) keeps the same form 
under the transformation $\theta \to -\theta$, 
if $\Psi(\theta)$ and $\Psi(\theta)^{*}$ are replaced by 
$\Psi(-\theta)^{*}$ and $\Psi(-\theta)$, respectively. 
This indicates that 
\bea
\Psi(-\theta)=\Psi(\theta)^{*} \quad {\rm and} \quad 
\Psi(-\theta)^{*}=\Psi(\theta) .
\label{psi-z2}
\eea
Furthermore, $\Omega$ is a real function, as shown in (\ref{eq:K3}). 
Using these properties, one can show that 
\begin{align}
\Omega(\theta)=&(\Omega(\theta))^*
              = \Omega(\Psi(\theta)^{*},\Psi(\theta),e^{-i\theta}) \notag\\
              =&\Omega(\Psi(-\theta),\Psi(-\theta)^{*}, e^{-i\theta})
              =\Omega(-\theta).
\end{align}
Thus, $\Omega$ is a periodic even function of $\theta$ with 
a period $2\pi/3$. 
The chiral condensate $\sigma(\theta)$ is also a 
periodic even function of $\theta$, 
$\sigma(\theta)=
\sigma(\theta+2 \pi k/3)=\sigma(-\theta)$, 
because it is given by $\sigma(\theta)=d\Omega(\theta)/dm_0$.

The modified Polyakov loop $\Psi$ has a periodicity of (\ref{psi-RW}). 
The real (imaginary) part of $\Psi$ is even (odd) under the interchange 
$\theta \leftrightarrow - \theta$, because of (\ref{psi-z2}): 
${\rm Re}[\Psi(\theta)]=(\Psi(\theta) + \Psi(\theta)^*)/2=
{\rm Re}[\Psi(-\theta)]$
and 
${\rm Im}[\Psi(\theta)]=(\Psi(\theta) - \Psi(\theta)^*)/(2i)
=-{\rm Im}[\Psi(-\theta)]$. 
Thus, the real (imaginary) part of $\Psi$ is a periodic even (odd) function of 
$\theta$.

Since $\Omega(\theta)$, $\Psi(\theta)$ and $\sigma(\theta)$ 
are periodic functions of 
$\theta$ with a period $2\pi/3$, here we think a period 
$0 \le \theta \le 2\pi/3$. In the region, 
periodic even functions such as 
$\Omega(\theta)$, $\sigma(\theta)$ and ${\rm Re}[\Psi(\theta)]$ are 
symmetric with respect to a line $\theta=\pi/3$. 
This indicates that such an even function has 
a cusp at $\theta=\pi/3$, 
if the gradient $\displaystyle \lim_{\theta \rightarrow \pi/3 \pm0} 
d\Omega/d\theta$ is neither zero or infinity. 
Such a cusp comes out in the high $T$ region, 
as shown later with numerical calculations. 
This means that the chiral phase 
transition at $\theta=\pi/3$ is the second order. 

Meanwhile, ${\rm Im}[\Psi(\theta)]$ is a periodic odd function, 
so that ${\rm Im}[\Psi(\pi/3-\epsilon)]=-{\rm Im}[\Psi(-\pi/3+\epsilon)]
=-{\rm Im}[\Psi(\pi/3+\epsilon)]$ for positive infinitesimal $\epsilon$.
This indicates that ${\rm Im}[\Psi(\theta)]$ 
is discontinuous at $\theta=\pi/3$, if it is not naught there. 
This is precisely the RW phase transition, and seen in the high $T$ region, 
as shown later. 
The deconfinement phase transition at $\theta=\pi/3$ is the 
first order transition appearing in the imaginary part of $\Psi$.

When $\Psi=\Psi^{*}=0$, $\Omega$ with $T$ fixed depends only on the baryon number chemical potential $\mu_{\rm B}$, and then not on the quark number chemical potential $\mu$ explicitly. 
In this sense, quarks are ``confined" in the PNJL model. 
Meanwhile, for the case of finite $\Psi$, $\Omega$ depends on both $\mu_{\rm B}$ and $\mu$, indicating that the system is in a mixed phase 
of baryons and quarks. 
Thus, the quark confinement is described by the PNJL model through the
percentage of $\mu_{\rm B}$ and $\mu$, and the order parameter $\Psi$ of
the ${\mathbb Z}_3$ symmetry  is found to play an pseudo order parameter
of the confinement transition practically.

Since the NJL model is nonrenormalizable, it is then needed to 
introduce a cutoff in the momentum integration. 
Here we take the three-dimensional momentum cutoff 
\begin{equation}
\int \frac{d^3{\bf p}}{(2 \pi)^3}\to 
{1\over{2\pi^2}} \int_0^\Lambda dp p^2.
\label{eq:E15}
\end{equation}
Hence, the present model has three parameters 
$m_0$, $\Lambda$, $G_{\rm s}$ in the NJL sector. 
Following Ref.~\cite{Kashiwa1}, we take $\Lambda =0.6315$ GeV and
$G_{\rm s}=5.498$ GeV$^{-2}$, although we consider the chiral 
($m_0=0$) limit.

\begin{figure}[htbp]
\begin{center}
 \includegraphics[width=0.4\textwidth]{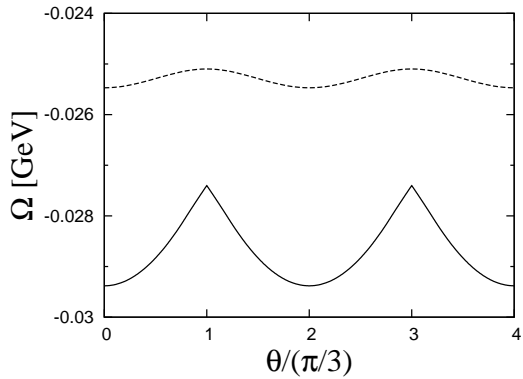} 
\end{center}
\caption{Thermodynamic potential $\Omega$ as a function of $\theta$. 
The solid line represents a result of the case of $T=300$~MeV, and 
the dashed one corresponds to that of $T=250$~MeV. 
}
\label{fig2}
\end{figure}

Figure \ref{fig2} shows $\Omega$ 
as a function of $\theta$ in two cases of 
$T=250$~MeV and 300~MeV. The potential $\Omega$ is smooth everywhere 
in the low $T$ case, but not at $\theta =(2k+1)\pi/{3}$ 
in the high $T$ case. 
This result is consistent with 
the RW prediction \cite{RW} and lattice simulation~\cite{Lomb}
on the $\theta$ and the $T$ dependence 
of the QCD thermodynamic potential. 

\begin{figure}[htbp]
\begin{center}
 \includegraphics[width=0.4\textwidth]{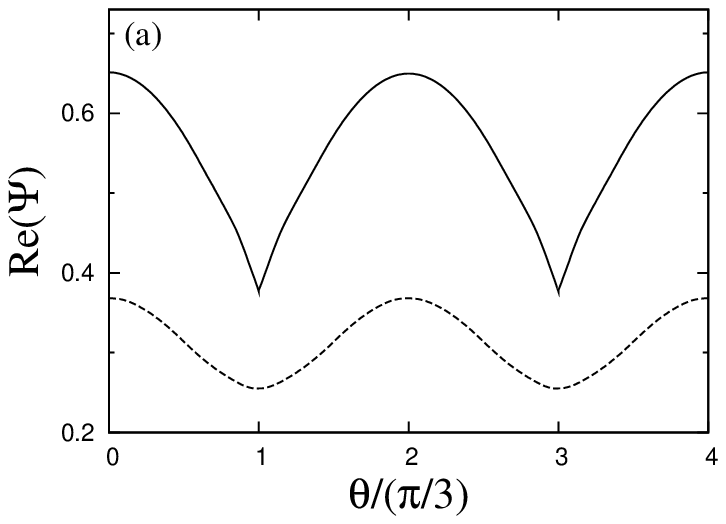} 
 \includegraphics[width=0.4\textwidth]{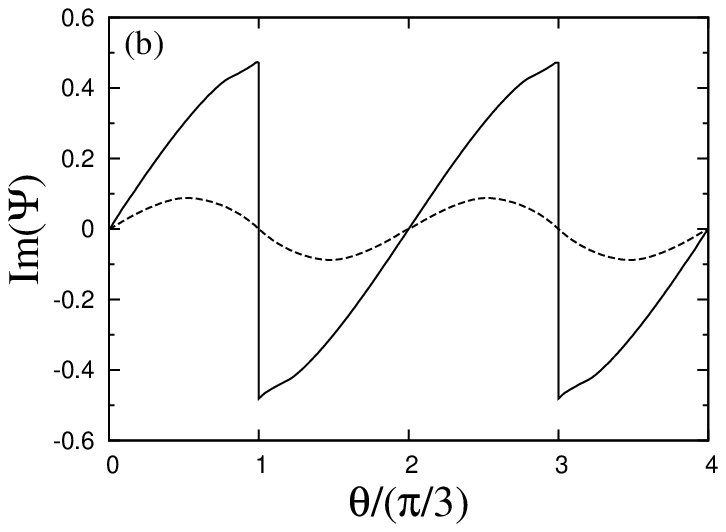} 
\end{center}
\caption{The modified Polyakov loop $\Psi(\theta)$ 
as a function of $\theta$; 
(a) for the real part and (b) for the imaginary part. 
The definitions of lines are the  same as in Fig.~2. }
\label{fig3}
\end{figure}

Figure \ref{fig3} shows the real and imaginary parts of 
the modified Polyakov loop $\Psi(\theta)$. 
In the case of $T=300$~MeV, the imaginary part of 
$\Psi(\theta)$ is discontinuous at $\theta =(2k+1)\pi/{3}$, 
while the real part of $\Psi(\theta)$ is continuous but 
not smooth there. 
Thus, the deconfinement phase transition of first order 
appears at $\theta =(2k+1)\pi/{3}$ in the high $T$ region. 
This is precisely the RW phase transition. 
In the case of $T=250$~MeV, meanwhile, both the real and the imaginary part 
are smooth everywhere. 
All the results on the $\theta$ and the $T$ dependence of $\Psi$ are 
consistent with the RW prediction on it and the 
results of lattice simulations~\cite{FP1,Elia,Chen}. 
The present analysis clearly shows that the transition is the first order in 
the imaginary part of $\Psi(\theta)$, while the preceding works 
discuss only the order of transition. 
Thus, the present analysis is more informative 
than the preceding analyses. 

\begin{figure}[htbp]
\begin{center}
 \includegraphics[width=0.4\textwidth]{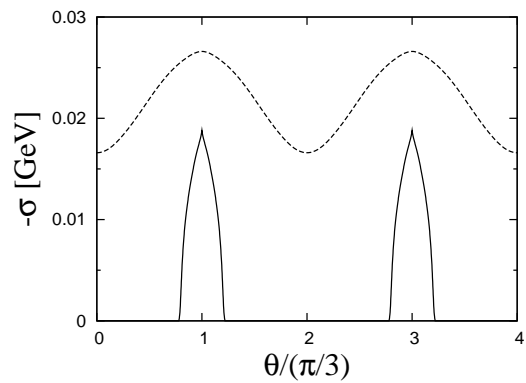} 
\end{center}
\caption{Chiral condensate $\sigma$ as a function of $\theta$.
The definitions of lines are the  same as in Fig.~2. }
\label{fig4}
\end{figure}

Figure \ref{fig4} 
shows the chiral condensate $\sigma$ as a function of $\theta$. 
In the case of $T=300$~MeV, $\sigma$ has a cusp at each of 
$\theta =(2k+1)\pi/{3}$. Thus, the chiral phase transition of second order 
comes out at $\theta =(2k+1)\pi/{3}$. 
Meanwhile, in the case of $T=250$MeV, 
there is no cusp at $\theta =(2k+1)\pi/{3}$, indicating no 
chiral phase transition there. 

\begin{figure}[htbp]
\begin{center}
 \includegraphics[width=0.4\textwidth]{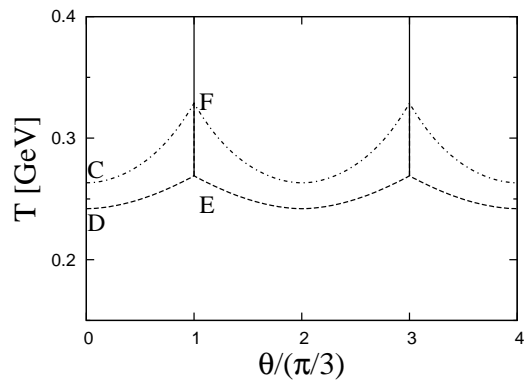} 
\end{center}
\caption{
The phase diagram in the $\theta$-$T$ plane. 
The solid vertical line starting from point 
E represents the RW deconfinement phase transition of first order. 
The dashed curve between D and E represents 
the deconfinement phase transition of crossover, and 
the dot-dashed curve between C and F does the second-order 
chiral phase transition. 
The second-order chiral phase transition also appears on the line between 
E and F. 
}
\label{fig5}
\end{figure}

Figure \ref{fig5} represents the phase diagram in the $\theta$-$T$ plane. 
The phase diagram is symmetric with respect to 
each of lines $\theta=k \pi/3$ for any integer $k$. 
The dashed curve between D and E represents 
the deconfinement phase transition of crossover, and 
the dot-dashed curve between C and F does the second-order 
chiral phase transition. 
For $\theta \neq k \pi/3$, thus, 
the chiral phase transition occurs at $T$ higher than 
the deconfinement phase transition does. 
The solid vertical line starting from point 
E represents the RW deconfinement phase transition of first order. 
Both the deconfinement and the chiral phase transition occur on 
the line between E and F, although 
the deconfinement phase transition is the first order and 
the chiral phase transition is the second order there. 
Point F turns out to be a bifurcation of 
the chiral phase transition line, and point E is the endpoint
of both the deconfinement and chiral phase transitions.

Temperatures of C, D, E, F are about 261~MeV, 240~MeV, 269~MeV, 328~MeV, 
respectively. Thus, at $\theta=0$ 
the critical temperature of the chiral phase transition is 
higher by about 20~MeV than that of the deconfinement transition, and 
the difference is getting larger gradually as $\theta$ increases to $\pi/3$. 
Meanwhile, the lattice simulation suggests that the two critical 
temperatures are almost identical 
no only for zero $\theta$ but also for finite $\theta$~\cite{Elia,Elia2}. 
The difference between the two critical temperatures is reduced 
by a factor 3 by 
adding the scalar-type eight-quark interaction 
to the PNJL Lagrangian \cite{Kashiwa3}. 
Further discussion will be made in the forthcoming paper.

In summary, 
the phase diagram in the $\theta$-$T$ plane is studied with 
the Polyakov loop extended Nambu--Jona-Lasinio 
(PNJL) model. Since the PNJL model possess 
an extended ${\mathbb Z}_{3}$ symmetry, 
quantities invariant under the symmetry, such as 
the thermodynamic potential, the chiral condensate and 
the modified Polyakov loop, automatically have 
the Roberge-Weiss periodicity that QCD does. 
The deconfinement phase transition of first order occurs 
at $\theta=(2k+1)\pi/3$ through the imaginary part of the modified 
Polyakov loop. This result is more informative than 
the RW prediction and the results of lattice QCD in 
which only the order of transition is discussed. 
The present model also clarifies the phase diagram of chiral transition 
in the chiral limit. In particular, it is of interest that 
there exists a bifurcation of the transition line. 
In this paper, our discussion is focused only on  
qualitative comparison with the results of 
lattice simulation. Quantitative comparison 
will be made in the forthcoming paper.

The success of the PNJL model comes from the fact 
that the PNJL model has the extended ${\mathbb Z}_{3}$ symmetry, more 
precisely that the thermodynamic potential (\ref{eq:K3}) is a function only of 
variables, $\Psi$, $\Psi^*$, $e^{\pm \beta\mu_{\rm B}}$ and $\sigma$, 
invariant under the extended ${\mathbb Z}_{3}$ symmetry. 
A reliable effective theory of QCD proposed in future 
is expected to have the same property 
in its thermodynamic potential. This may be a good guiding principle 
to elaborate an effective theory of QCD.

\noindent
\begin{acknowledgments}
The authors thank M. Matsuzaki and T. Murase for useful discussions 
and suggestions. 
H.K. thanks M. Imachi, H. Yoneyama and M. Tachibana 
for useful discussions about the RW phase transition. 
This work has been supported in part 
by the Grants-in-Aid for Scientific Research 
(18540280) of Education, Science, Sports, and Culture of Japan.
\end{acknowledgments}


\end{document}